\documentclass[runningheads,a4paper]{llncs}

\title{TREEOME: A framework for epigenetic and transcriptomic data integration to explore regulatory interactions controlling transcription}
\author{David M. Budden\,$^{1,2}$, Daniel G. Hurley\,$^{1}$ and Edmund J. Crampin\,$^{1,2,3,4,5}$}

\institute{$^{1}$Systems Biology Laboratory, Melbourne School of Engineering,
$^{2}$NICTA Victoria Research Laboratory,
$^{3}$ARC Centre of Excellence in Convergent Bio-Nano Science and Technology,
$^{4}$Department of Mathematics and Statistics and $^{5}$School of Medicine,
The University of Melbourne, Parkville, Victoria 3010, Australia\\
\texttt{\{david.budden, daniel.hurley, edmund.crampin\}@unimelb.edu.au}}

\date{}

\usepackage[table]{xcolor}
\usepackage{times}
\usepackage{url}

\usepackage{graphicx}
\usepackage{amsmath}

\usepackage{epstopdf}

\begin{document}
\maketitle

\begin{abstract}

\ \\ \textbf{Motivation:} Predictive modelling of gene expression is a powerful framework for the \emph{in silico} exploration of transcriptional regulatory interactions through the integration of high-throughput -omics data. A major limitation of previous approaches is their inability to handle conditional and synergistic interactions that emerge when collectively analysing genes subject to different regulatory mechanisms. This limitation reduces overall predictive power and thus the reliability of downstream biological inference.



\ \\ \textbf{Results:} We introduce an analytical modelling framework (TREEOME: tree of models of expression) that integrates epigenetic and transcriptomic data by separating genes into putative regulatory classes. Current predictive modelling approaches have found both DNA methylation and histone modification epigenetic data to provide little or no improvement in accuracy of prediction of transcript abundance despite, for example, distinct anti-correlation between mRNA levels and promoter-localised DNA methylation. To improve on this, in TREEOME we evaluate four possible methods of formulating gene-level DNA methylation metrics, which provide a foundation for identifying gene-level methylation events and subsequent differential analysis, whereas most previous techniques operate at the level of individual CpG dinucleotides. We demonstrate TREEOME by integrating gene-level DNA methylation (bisulfite-seq) and histone modification (ChIP-seq) data to accurately predict genome-wide mRNA transcript abundance (RNA-seq) for H1-hESC and GM12878 cell lines.

\ \\ \textbf{Availability:} TREEOME is implemented using open-source software and made available as a pre-configured bootable reference environment. All scripts and data presented in this study are available online at\\ \texttt{http://sourceforge.net/projects/budden2015treeome/}.

\ \\ \textbf{Contact:} \texttt{edmund.crampin@unimelb.edu.au}


\end{abstract}

\section{Introduction}
\label{sec:introduction}

Understanding the precise spatiotemporal regulation of eukaryotic gene expression is a central challenge in molecular biology. Transcriptional regulation is governed by dynamic restructuring of chromatin to control gene accessibility, mediated by post-translational modifications of nucleosomal histone proteins. Any perturbation of the systems regulating gene accessibility can affect critical cellular functions including homeostasis, differentiation and apoptosis. Consequently, dysregulation of these systems has been implicated with hundreds of developmental, autoimmune, neurological, inflammatory and neoplastic disorders~\cite{portela2010epigenetic}.

The relationship between histone modifications and gene expression involves complex systems of protein-mediated regulatory events that are still poorly understood. The simplest interactions involve acetlyation of lysine residues on the histone H3/4 amino-termini, reducing their net-positive charge and weakening charge-dependent interactions with adjacent nucleosomes and the negatively-charged DNA backbone~\cite{bannister2011regulation}. Promoter-localised histone acetylation is thus considered a euchromatic modification, as it promotes the establishment of open DNase-sensitive chromatin and active transcription. Histone lysine methylation is further separated from the physical transcription process; mono-, di- and tri-methylation of specific residues are recognised by proteins with varying and context-sensitive regulatory roles, including the Polycomb repressive complexes (associated with H3K27me2/3) and DNA \emph{de novo} methyltransferase family (associated with H3K9me2/3).

 To develop a comprehensive understanding of the regulatory logic controlling eukaryotic gene expression by studying individual protein-protein interactions would require a currently-unavailable volume and resolution of proteomics data. Instead, predictive modelling frameworks have been developed that leverage the wealth of high-throughput sequencing data generated by recent large-scale consortia (\emph{e.g.}~\cite{encode2012integrated}) to study the indirect relationships between histone modifications and transcript abundance. The utility of these models is not only the ability to predict RNA abundance for individual species (at which the best models currently available perform rather poorly at the level of individual genes), but rather the biological insights that can be gained by exploring the relationships inferred from the data; the prediction accuracy of such models is simply an indirect measure of their explorative potential.

We have previously reviewed predictive modelling in the context of eukaryotic transcriptional regulation~\cite{budden2014predictive}, which has been applied to a wide range of problems in molecular biology. These include: inferring regulatory roles of transcription factors from their respective binding motifs~\cite{mcleay2012genome}; identifying regulatory elements responsible for differential expression patterns~\cite{cheng2012modeling}; exploring the relationship between gene expression and higher-order chromatin domains~\cite{budden2014exploring}; and large-scale comparative analysis of the transcriptome across distant species~\cite{gerstein2014comparative}.

\begin{figure}[t]
\begin{center}
    \includegraphics[width=0.48\textwidth]{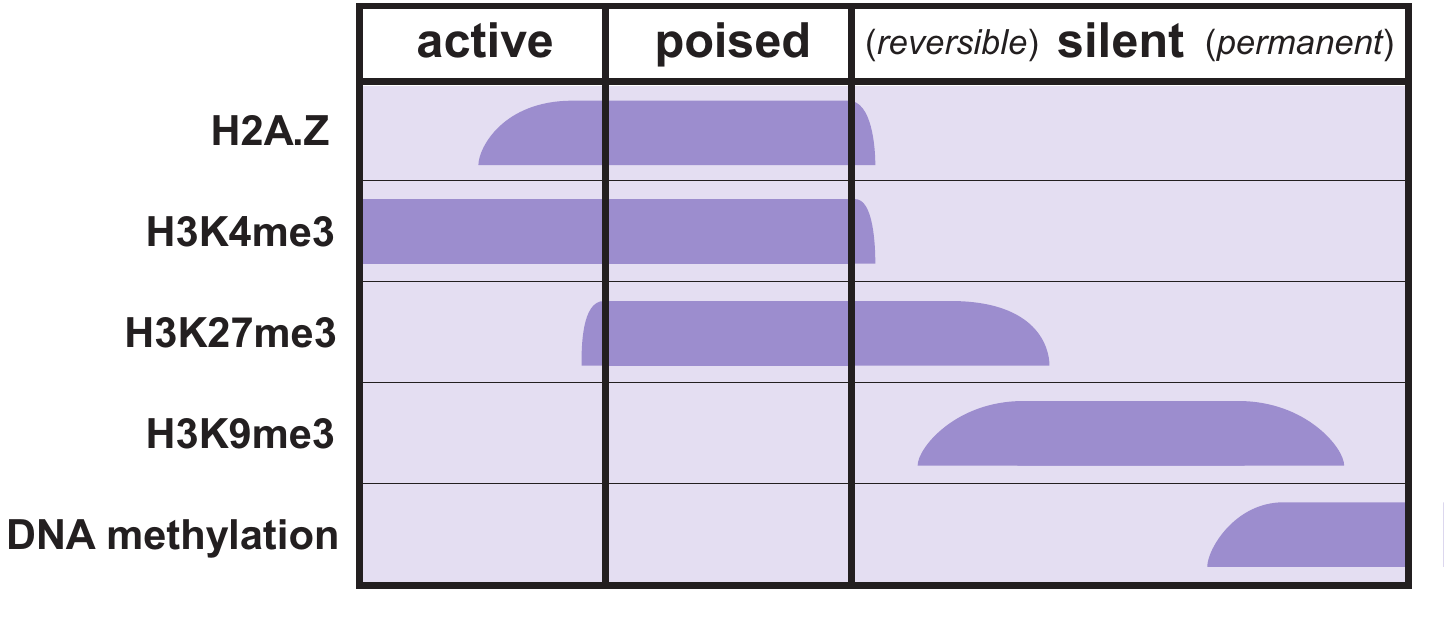}
    \end{center}
    \caption{Illustration of the histone/epigenetic code in the context of the promoter-localised regulatory elements analysed in this study. Only active genes exhibit significant expression, corresponding with H3K4me3 often flanked by H2A.Z. Poised and reversible/permanently silenced genes are distinguished by decreasing likelihood of genes returning to an active state; poised genes are marked by bivalent H3K4/27me3 and H2A.Z, while silent genes are marked by H3K27me3 (facultative heterochromatin), H3K9me3 (constitutive heterochromatin) and DNA methylation (permanent silencing).}
    \label{fig:histonecode}
\end{figure}

Despite the utility of predictive modelling as a framework for uncovering novel molecular biology, a major limitation of current approaches is their inability to handle conditional and synergistic interactions that emerge when analysing genes subject to different regulatory mechanisms. For this study, we have selected four histone modifications that epitomise this problem through their indirect and context-sensitive regulatory roles: H3K4me3, H3K27me3, H3K9me3 and the H2A.Z histone variant. A simplified histone/epigenetic code for these modifications is illustrated in Figure~\ref{fig:histonecode}.

As an example of regulatory heterogeneity, Figure~\ref{fig:histonecode} illustrates that H3K4me3 (commonly associated with gene activation~\cite{bannister2011regulation}) promotes transcription in the absence of H2A.Z, but in the presence of H2A.Z only promotes transcription if H3K27me3 is absent~\cite{voigt2013double}. This description of conditional behaviour remains an oversimplification of the underlying regulatory events, as the transcriptional effect of histone lysine methylation depends on both the level of methylation (mono/di/tri), location within the gene (5' versus 3') and the presence of other regulatory elements (\emph{e.g.} transcription factors and ncRNAs). Current predictive modelling approaches generally assume linear/additive models and are unable to capture these conditional relationships. Although some studies have investigated the application of non-linear regression models (\emph{e.g.} support vector regression~\cite{cheng2011statistical}), quantitative analysis of these models across multiple scenarios has revealed that they perform no better than standard linear models~\cite{budden2014predictive}. We speculate that this analytical limitation is a major cause of the distinct lack of models integrating DNA methylation data in previous studies, as low-expression genes may be under the control of a variety of other silencing mechanisms.

In this study, we introduce an analytical framework (TREEOME: tree of models of expression) that facilitates the integration of regulatory data by separating genes into putative regulatory classes on the basis of histone modification and/or DNA methylation state. We demonstrate TREEOME by integrating gene-level DNA methylation (bisulfite-seq) and histone modification (ChIP-seq) data to predict genome-wide RNA transcript abundance (RNA-seq) for H1-hESC and GM12878 cell lines. As there has been little previous work in formulating and/or evaluating gene-level DNA methylation statistics, our analysis is prefaced by a quantitative evaluation of four possible promoter-localised methylation scores. These methylation scores provide a foundation for identifying significant methylation and subsequent differential analysis, at the level of genes rather than individual CpG dinucleotides.

\section{Methods and materials}

\subsection{Gene-specific histone scores}

The association strength between each gene, $i$, and histone modification, $j$, is calculated using the constrained sum-of-tags histone score~\cite{budden2014predictive}:

\begin{equation}
\label{eq:histonescore}
a_{ij} = \sum_kg_k,
\end{equation}

\noindent{where $g_k$ is the is the number of ChIP-seq reads (or normalised equivalent) for $j$ mapped to position $k$ relative to the TSS of $i$. As ChIP-seq involves sequencing of DNA corresponding with the end of each nucleosome, the position for each read was shifted by $\pm 73$ bp (for $\pm$ strand respectively) to centre on the modified nucleosome~\cite{karlic2010histone}. Integrating over a region 2000 bp either side of the TSS (approximating the average width of histone modification ChIP-seq binding regions) is standard for this approach~\cite{cheng2011statistical,cheng2012modeling,mcleay2012genome} and applied throughout this study.

\subsection{Predictive modelling of gene expression}
\label{ss:predictive}


In this study, we model the RPKM-normalised transcript abundance, $y_i$, of each gene, $i$, as a general linear function of its association, $a_{ij}$, with each histone modification, $j$:

\begin{equation}
\label{eq:regressionmodel}
\sinh^{-1}(y_i) = \mu + \sum_j\beta_ja_{ij} + \varepsilon_i,
\end{equation}

\noindent{where $\beta_j$ captures the influence of histone modification $j$ on gene expression, $\mu$ is the basal expression level, and $\varepsilon_i$ is the gene-specific error term. The inverse hyperbolic sine (arsinh) transformation, $\sinh^{-1}(x) = \log(x + \sqrt{1 + x^2})$, is approximately equal to $\log(2x)$ for $x \gg 0$, allowing it to be regarded as practically-equivalent to the log-transformation applied in previous gene expression modelling studies~\cite{budden2014predictive}. Unlike $\log(x)$, $\sinh^{-1}(x)$ is defined for $x=0$, removing the need to meta-optimise small constants to add to $x$ (leading to spurious inflation of prediction accuracy) and making it better-suited to integrating ChIP-seq and RPKM-normalised RNA-seq data.}

\subsection{Modelling conditional regulatory interactions with decision trees}
\label{ss:decisiontrees}



The decision tree framework (TREEOME) mitigates the analytical consequences of conditional and synergistic interactions in gene expression data. For example, gene-level H2A.Z scores (an indicator of histone bivalency) could be used to separate genes into two subsets: those putatively-regulated by H2A.Z and those that are not. Separate predictive models can then be constructed and evaluated for both subsets from the remaining regulatory elements, as illustrated in Figure~\ref{fig:flowchart1}.

\begin{figure}[t]
\begin{center}
    \includegraphics[width=0.48\textwidth]{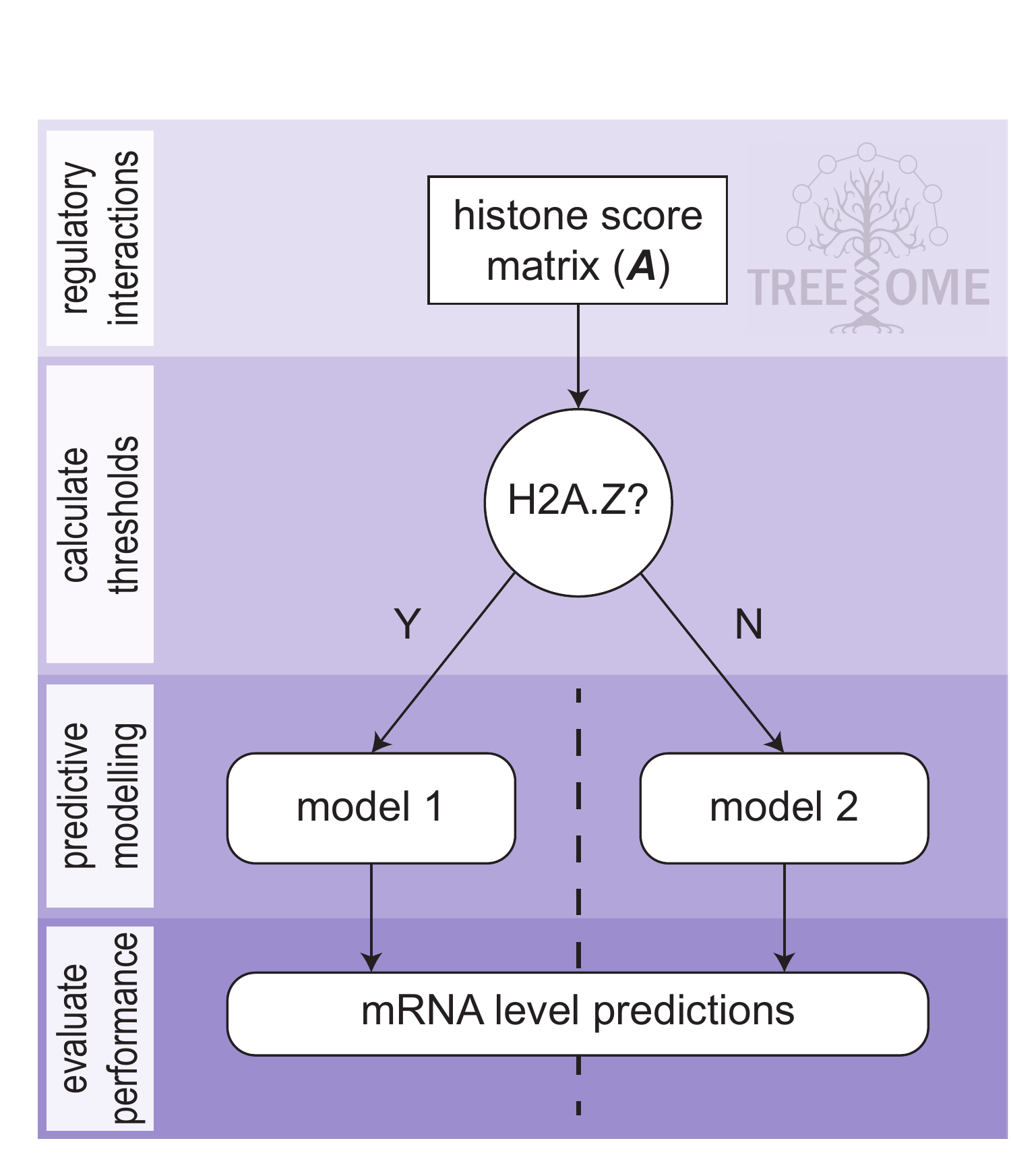}
    \end{center}
    \caption{Illustration of a general predictive modelling pipeline where the H2A.Z histone variant has been used to separate genes into two categories. Categorising genes by the presence of promoter-localised H2A.Z removes significant heterogeneity in the regulatory role of H3K4me3; H3K4me3 in the presence of H2A.Z is often a hallmark of low expression (\emph{i.e.} poised genes), whereas H3K4me3 is otherwise associated with active transcription. These synergistic interactions are poorly-handled by current regression modelling.}
    \label{fig:flowchart1}
\end{figure}


TREEOME uses an unsupervised method to define the threshold above which a gene-level histone score is accepted to represent actual regulatory activity:
\begin{itemize}
\item Under the assumption that H2A.Z is sufficient to separate genes into two sets of homogeneously-regulated genes (or where both sets are to be further subdivided according to other epigenetic markers), the threshold is chosen to maximise the combined prediction accuracy of both models.
\item Under the assumption that only one subset is homogeneously-regulated (\emph{i.e.} the other is to be further subdivided), the threshold is chosen to maximise the prediction accuracy of the homogenous model.
\end{itemize}

\noindent{TREEOME implements a greedy algorithm that is not necessarily globally optimal. Although improved prediction accuracy could be obtained by optimising over the full set of thresholds for an arbitrarily-large tree, this approach would lose the biological meaning (regulated-or-not) underlying our threshold selection methodology.}

\subsection{Evaluation of prediction accuracy}
\label{ss:evaluation}


Prediction accuracy is assessed for each regression model using an adjusted $R^2$ score, which in comparison to the standard $R^2$ approach prevents spurious inflation of the statistic due to introduction of additional explanatory variables~\cite{harel2009estimation}.

Separate RNA-seq replicates were used for model training and evaluation to prevent over-fitting to experimental noise. If multiple replicates are not available, the adjusted $R^2$ score for each model can be determined using a $k$-fold cross-validation process. 

\subsection{Derivation of putative regulatory roles}
\label{ss:pca}


Putative regulatory roles are inferred for each histone modification using principal component analysis (PCA). Specifically, the histone score matrix, $\mathbf{A}$ (see eq. (\ref{eq:histonescore})), for a gene-set of interest is arsinh-transformed and reformulated using the following singular value decomposition~\cite{watkins2004fundamentals}:

\begin{equation}
\label{eq:pca}
\sinh^{-1}(\mathbf{A}) = \mathbf{U}\boldsymbol{\Sigma} \mathbf{V}^{\top},
\end{equation}

\noindent{where $\mathbf{U}$ is the matrix of component scores, $\boldsymbol{\Sigma}$ is the diagonal matrix of the singular values of $\mathbf{A}$, and $\mathbf{V}$ is the matrix of loadings (weights by which the histone scores are multiplied to derive their respective component scores). In the context of modelling gene expression, the columns of the matrix $\mathbf{U}\boldsymbol{\Sigma}$ are the principal components (PCs), and the rows correspond with eigengenes~\cite{alter2000singular}. The data-derived, putative regulatory role of each histone modification is simply its contribution (loading) toward the individual PC most predictive of gene expression~\cite{budden2014predictive}.}

\subsection{Quantifying gene-level DNA methylation}
\label{ss:methscore}


Compared to CpG-level methylation scores, gene/region-level DNA methylation scores are not well-established in previous literature. We explore four possible promoter-localised scores in the context of predictive gene expression modelling, considering a window 2000 bp either side of the respective gene's TSS:

\begin{itemize}
\item \textbf{Sum of methylation fractions by site (SMFS)}: Sum of the CpG-level methylation scores within a region, similar to the constrained sum-of-tags score previously applied to the analysis of ChIP-seq data~\cite{budden2014predictive}
\item \textbf{Mean methylation fraction by site (MMFS)}: Equivalent to the SMFS score divided by the number of assayed CpGs within the region, similar to the mean methylation level described by~\cite{schultz2012leveling}
\item \textbf{Mean methylation fraction by region (MMFR)}: Proportion of raw reads that were found to be methylated, similar to the weighted methylation level described by by~\cite{schultz2012leveling}
\item \textbf{Sum of scaled methylation reads by region (SMRR)}: Equivalent to the MMFR score where each read is multiplied by $-\exp(d/d0)$, where $d$ is the distance (bp) from the TSS and $d0 = 5000$, similar to the exponentially decaying affinity score previously applied to the analysis of ChIP-seq data~\cite{budden2014predictive}
\end{itemize}

\subsection{Data}

H1-hESC and GM12878 data were selected to demonstrate TREEOME for both pluripotent and differentiated cell lines, as functional patterns of DNA methylation vary significantly across the lineage commitment spectra. All H1-hESC and GM12878 gene expression (RNA-seq), histone modification (ChIP-seq) and DNA methylation (methyl RRBS) data were downloaded from ENCODE~\cite{encode2012integrated}. Specific GEO accession numbers for each dataset are provided in Table~\ref{tab:data1}. The TSS for each gene was taken from the gene annotation dataset for the human genome (hg19/GRCh37). Multiple transcripts or isoforms were removed by considering only the most 5'-located TSS for each unique Ensembl gene identifier, resulting in a set of 11,806 genes for analysis. RNA-seq data was re-mapped to hg19 using Subread~\cite{liao2013subread} and RPKM-normalised using edgeR~\cite{mortazavi2008mapping,robinson2010edger}.

\begin{table}
\begin{center}
\renewcommand{\arraystretch}{1.3}
\caption{All H1-hESC and GM12878 data used in this study~\cite{encode2012integrated}.\label{tab:data1}}
\begin{tabular}{|p{2.5cm} | p{5cm}|}
\hline
\multicolumn{1}{|c|}{\textbf{Data type}} & \multicolumn{1}{c|}{\textbf{Data source}} \\
\hline
RNA-seq & GSM958730 (GM12878, 2 replicates)\\
& GSM958737 (H1-hESC, 2 replicates)\\
\hline
TSS & Ensembl hg19/GRCh37\\
&~\cite{flicek2013ensembl}\\
\hline
Methyl RRBS & GSM683906 (replicate 1)\\
(GM12878)& GSM683927 (replicate 2)\\
\hline
ChIP-seq & GSM733767 (H2A.Z)\\
(GM12878)& GSM733758 (H3K27me3)\\
& GSM733708 (H3K4me3)\\
& GSM733664 (H3K9me3)\\
\hline
Methyl RRBS & GSM683770 (replicate 1)\\
(H1-hESC)& GSM683879 (replicate 2)\\
\hline
ChIP-seq & GSM1003579 (H2A.Z)\\
(H1-hESC)& GSM733748 (H3K27me3)\\
& GSM733657 (H3K4me3)\\
& GSM1003585 (H3K9me3)\\
\hline
\end{tabular}
\end{center}
\end{table}

\subsection{Implementation}

TREEOME is implemented using open-source software and made available as a pre-configured bootable virtual environment using the approach described by~\cite{hurley2014virtual}. This environment was created using a minimal installation of Lubuntu 13.10; a lightweight Linux distribution which supports all the tools required. R version 3.0.1 was installed, along with the core set of packages and utilities required to explore the presented results. Alternatively, all data and scripts are available online at\\ \texttt{http://sourceforge.net/projects/budden2015treeome/}.

\section{Results and discussion}

We used TREEOME to build models based on dividing H1-hESC and GM12878 data into gene sets for regression modelling on the basis of DNA methylation score and subsequently (for genes identified as not methylated) on the basis of H2A.Z, a well-studied indicator of histone bivalency and poised expression~\cite{voigt2013double}. First, we establish which combination of gene-level histone and DNA methylation scores are most appropriate for our subsequent analyses, as described below.

\subsection{Histone modifications are predictive of transcript abundance}
\label{ss:results1}

To validate whether our histone score (eq. \ref{eq:histonescore}) and regression model (eq. \ref{eq:regressionmodel}) formulations are suitable for the data considered in this study, we evaluated the accuracy of model-predicted RPKM-normalised transcript abundance compared to actual RNA-seq data genome-wide for each cell line. These results are presented in Figure~\ref{fig:figure1}, and the performance of our models (adjusted $R^2$ = 0.43 for H1-hESC and 0.47 for GM12878) were found to be similar to those of previous studies~\cite{cheng2011statistical,cheng2012modeling,karlic2010histone,mcleay2012genome}.

Figure~\ref{fig:figure1} also presents the distribution of gene expression levels and data-derived putative regulatory roles of each histone modification genome-wide (methodology described in Sec.~\ref{ss:pca}), with positive/negative loadings suggesting activator/respressor roles respectively. It is evident that the differentiated lymphoblastoid GM12878 cell line exhibits more near-zero expression (silenced) genes than pluripotent H1-hESC, as expected due to DNA methylation-mediated gene silencing during lineage commitment. DNA methylation is further implicated by the stronger regulatory signal for H3K9me3 in GM12878, which is associated with DNA \emph{de novo} methyltransferase activity~\cite{cedar2009linking}.

\begin{figure*}[t]
\begin{center}
    \includegraphics[width=0.8\textwidth]{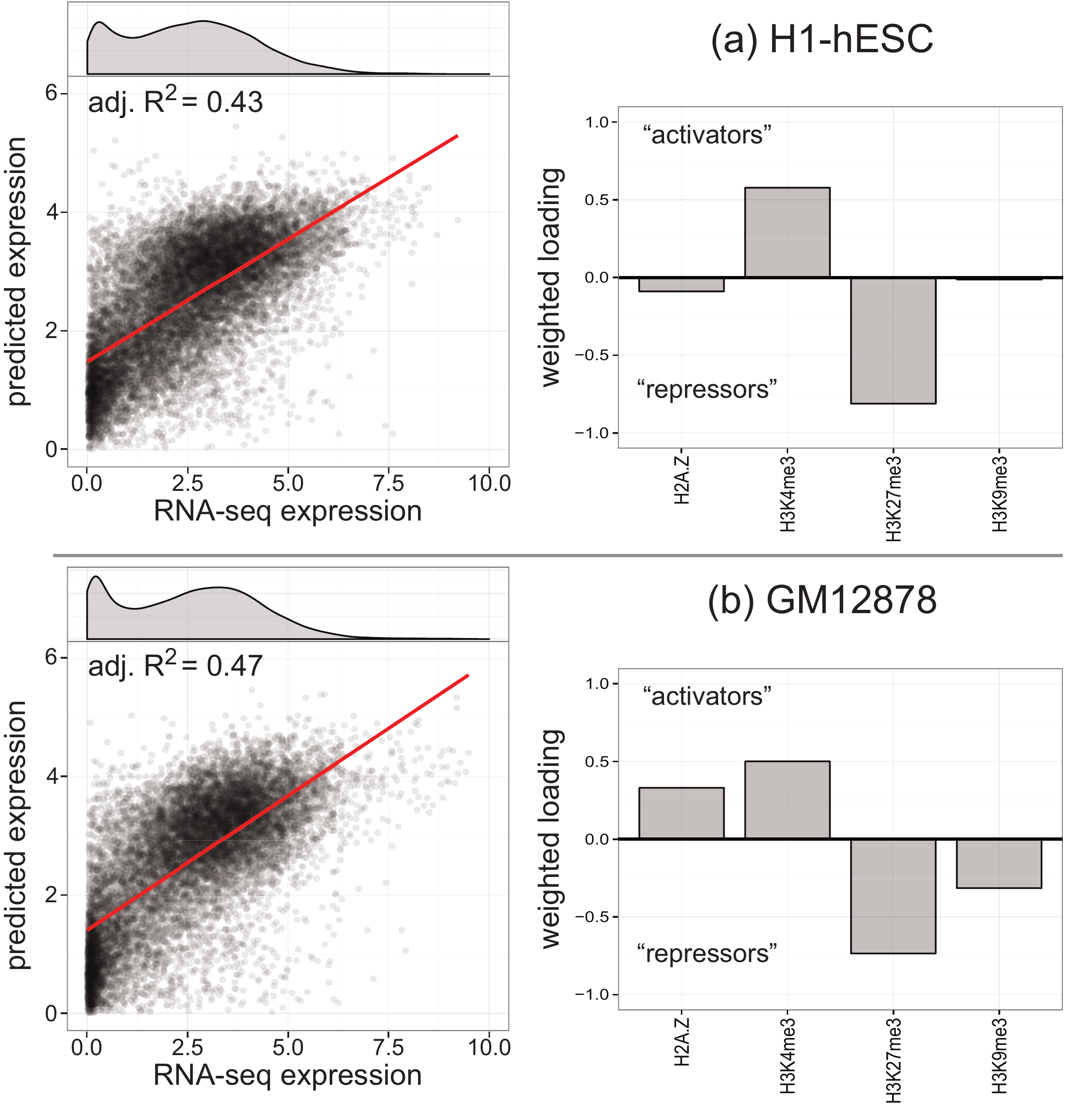}
    \end{center}
    \caption{Analysis of predictive models of genome-wide transcript abundance for (a) H1-hESC and (b) GM12878 cell lines, constructed from H2A.Z, H3K4me3, H3K27me3 and H3K9me3 histone scores. Both panels demonstrate the following: (top-left) the distribution of arsinh-transformed RPKM-normalised transcript abundance derived from RNA-seq data; (left) predicted-versus-measured transcript abundance for the linear regression model, with performance quantified as an adjusted $R^2$ score; and (right) the data-derived putative regulatory roles of each histone modification, with positive/negative loadings suggesting activator/respressor roles respectively. Of particular interest is the latent signature of DNA methylation-mediated gene silencing, with GM12878 exhibiting a higher proportion of near-zero expression genes and strikingly stronger regulatory signal for H3K9me3 (implicated in DNA \emph{de novo} methyltransferase activity), as expected following lineage-commitment.}
    \label{fig:figure1}
\end{figure*}

\subsection{MMFS is the most informative methylation score}
\label{ss:results2}

It is widely accepted that promoter-localised CpG methylation prevents the initiation of eukaryotic gene transcription~\cite{jones2012functions}. By extension, a suitable gene-level DNA methylation score should be anti-correlated with transcript abundance derived from genome-wide RNA-seq data. Figure~\ref{fig:figure2} presents the correlation between transcript abundance and each of the four DNA methylation scores described in Section~\ref{ss:methscore} (SMFS, MMFS, MMFR and SMRR) for all replicate combinations. MMFS performed equal-best for H1-hESC (Pearson's $r = -0.25$) and outright best for GM12878 (Pearson's $r = -0.31$), with all scores exhibiting stronger anti-correlation with GM12878 than hESC, as expected from Section~\ref{ss:results1}.

The distribution of promoter methylation (MMFS) versus transcript abundance presented in Figure~\ref{fig:figure2} demonstrates two distinct clusters, corresponding with active/unmethylated (green) and silenced/methylated genes (red). It is also evident that a large number of genes exhibit near-zero expression despite a lack of substantial DNA methylation (blue); these genes reduce the predictive power of DNA methylation in a current modelling framework and are likely silenced by other mechanisms (\emph{e.g.} repressor/silencer transcription factors~\cite{spitz2012transcription} or H3K27me3-mediated Polycomb activity ~\cite{cedar2009linking}).

\begin{figure*}[t]
\begin{center}
    \includegraphics[width=0.9\textwidth]{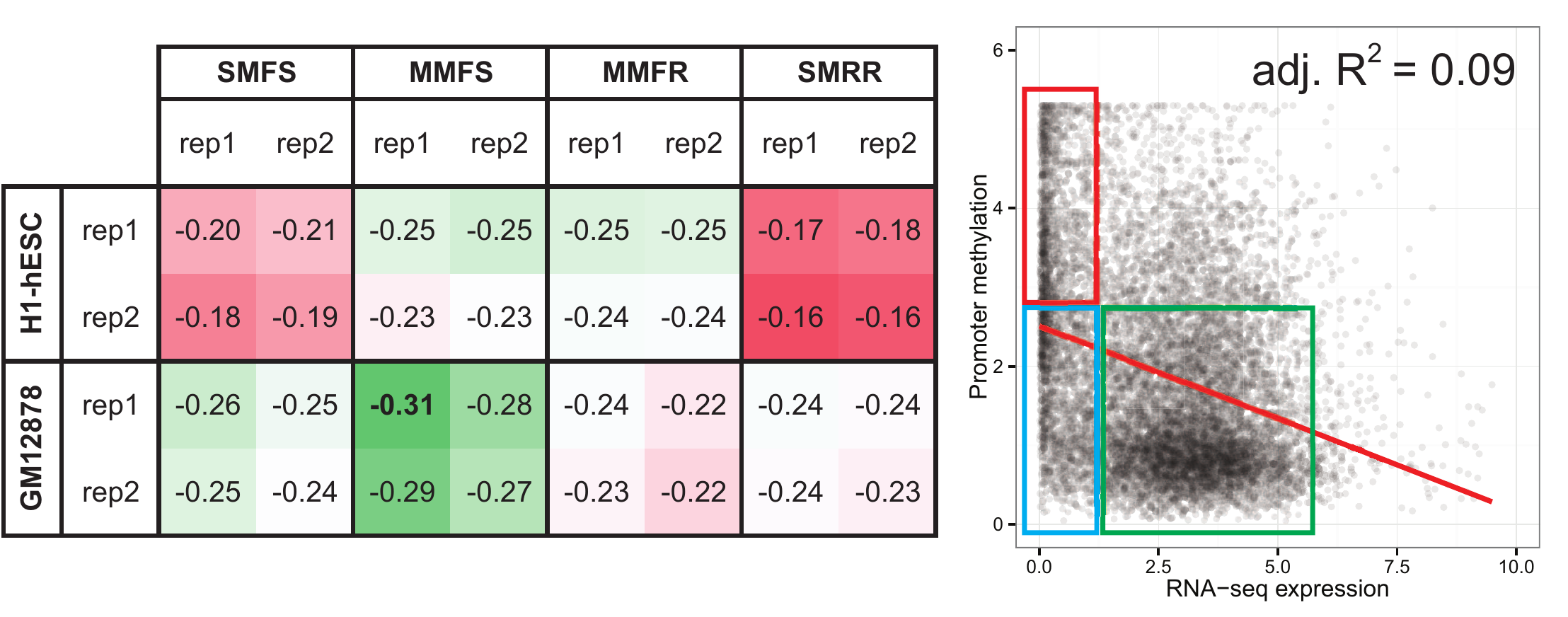}
    \end{center}
    \caption{Analysis of the gene-level DNA methylation scores described in Section~\ref{ss:methscore} (SMFS, MMFS, MMFR and SMRR). (Left) MMFS exhibits the strongest overall anti-correlation with RPKM-normalised transcript abundance (Pearson's $r = -0.31$), indicating that it is most appropriate for capturing the gene silencing effect of promoter-localised methylation. Model performance colour-coded by correlation, with the best/worst-performing models highlighted in green/red respectively. (Right) promoter methylation (MMFS) versus transcript abundance genome-wide for GM12878 (regression line shown in red), demonstrating two distinct gene clusters: active/unmethylated (green) and silent/methylated (red). It is also evident that a large number of genes exhibit near-zero expression despite a lack of substantial DNA methylation (blue); these genes reduce the predictive power of DNA methylation genome-wide and are likely silenced by other mechanisms (\emph{e.g.} repressor/silencer transcription factors~\cite{spitz2012transcription} or H3K27me3-mediated Polycomb activity~\cite{cedar2009linking}).}
    \label{fig:figure2}
\end{figure*}

\subsection{Na\"ive predictive model integration is unsuitable for DNA methylation data}
\label{ss:results3}

As demonstrated in Figure~\ref{fig:figure2}, all four gene-level DNA methylation scores are anti-correlated with genome-wide RNA transcript abundance, as expected due to the well-established silencing role of promoter-localised CpG methylation~\cite{jones2012functions}. Intuitively, integrating any of these scores into a gene expression model (particularly MMFS) should yield improved prediction accuracy due to the addition of information regarding methylation-mediated silencing.

A na\"ive approach to integrating DNA methylation into the predictive modelling framework (described in Section~\ref{ss:predictive}) involves simply concatenating the vector of methylation scores as a new column of the $n\times m$ histone score matrix, $\mathbf{A}$, where $n$ is the number of genes and $m$ is the number of histone modifications. We constructed these models for all combinations of cell line and DNA methylation score and found that the resultant change in prediction accuracy was negligible in all cases ($|\Delta\mathrm{adj.} R^2| < 10^{-3}$).



Strikingly, despite the anti-correlation shown between each methylation score and RNA transcript abundance, the na\"ive integration of this information into predictive models trained on histone modification data yields practically-zero improvement in prediction accuracy (irrespective of score or cell line). Within the constraints of a linear regression framework, DNA methylation and the four considered histone modifications are statistically redundant with respect to gene expression (similar redundancy between histone modifications and transcription factors has recently been explored in detail by~\cite{budden2014exploring}). These results indicate that a more principled approach of integrating transcriptional regulatory data is necessary to better leverage biological insight from predictive models.

\subsection{TREEOME improves predictive power for homogeneous regulatory classes}

We use the best-performing methylation score (MMFS) to separate genes into two putative regulatory classes (MMFS$^+$ versus MMFS$^-$). Intuitively, this approach should isolate genes subject to H3K9me3/DNA methylation-mediated silencing from an otherwise-heterogeneous set.

Unmethylated genes are still subject to a variety of transcriptional regulatory mechanisms, including H3K4me3-mediated euchromatinisation (activation) and H3K27me3-mediated facultative heterochromatinisation (repression)~\cite{li2007role}. As described in Section~\ref{ss:decisiontrees}, our ability to identify the signatures of these mechanisms is confounded by bivalency, where the otherwise antagonistic H3K4/27me3 are maintained in metastable equilibrium by the H2A.Z histone variant~\cite{voigt2013double}. Therefore, to further remove synergistic effects from our predictive models, the aforementioned set of MMFS$^-$ genes is separated into two further putative regulatory classes by H2A.Z score (H2A.Z$^+$ and H2A.Z$^-$). The final decision tree structure is illustrated in Figure~\ref{fig:tree_results1}.

\begin{figure*}
\begin{center}
    \includegraphics[width=0.9\textwidth]{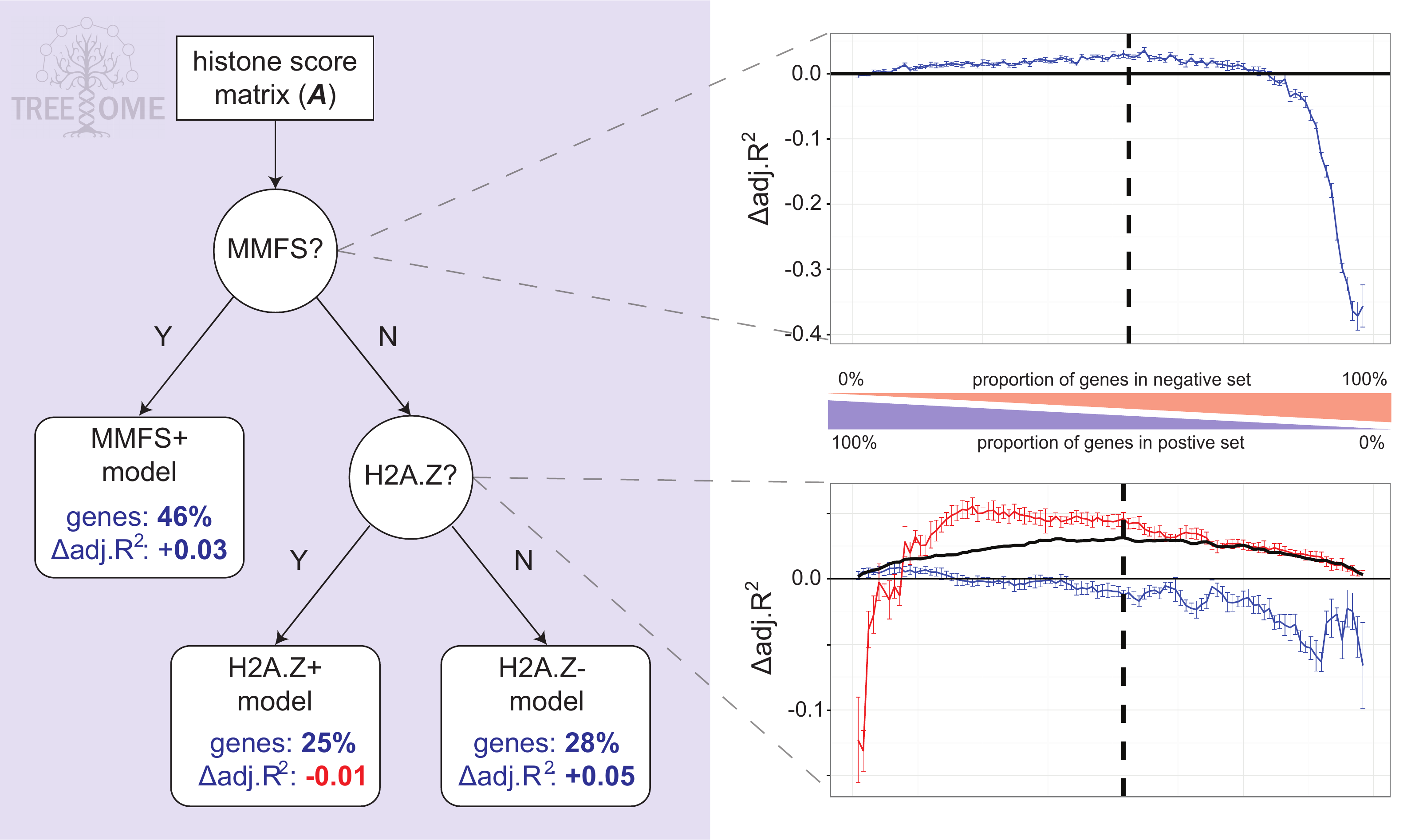}
    \end{center}
    \caption{Decision tree of predictive models of H1-hESC RNA transcript abundance, constructed from the same data as the model described in Section~\ref{ss:results3}. This tree uses promoter-localised DNA methylation (MMFS) and the H2A.Z histone variant to classify genes into three putative regulatory classes: MMFS$^+$ (high MMFS score), H2A.Z$^+$ (low MMFS and high H2A.Z) and H2A.Z$^-$ (low MMFS and low H2A.Z). A fourth category (high/high) would be biologically meaningless as DNA methylation and H2A.Z are mutually exclusive \emph{in vivo}~\cite{zilberman2008histone}. Thresholds were learned directly from the data using the unsupervised approach described in Section~\ref{ss:decisiontrees}. Specifically, the blue, red and black lines illustrate $\Delta\mathrm{adj.}R^2$ (relative to a standard model constructed from the same data) as a function of threshold values for positive (\emph{e.g.} high MMFS), negative and cumulative models respectively, with the optimal value for both forks indicated by a black dashed line. Error bars capture the standard error of the mean ($\mu\approx0$) for models constructed from 100 randomly-sampled gene-sets of equal size, illustrating the performance variation expected by chance (\emph{i.e.} fewer genes equals larger variation in model performance, as expected).}
    \label{fig:tree_results1}
\end{figure*}

In addition to the decision tree structure, Figure~\ref{fig:tree_results1} demonstrates the following for the H1-hESC cell line: the threshold selection process (described in Section~\ref{ss:decisiontrees}); the proportion of genes attributed to each putative regulatory class; and the respective performance results ($\Delta\mathrm{adj.}R^2$) relative to an unseparated regression model constructed from the same data. The statistics for both H1-hESC and GM12878 TREEOME analyses are presented in Table~\ref{tab:tree1results}.

\begin{table}[t]
\setlength{\tabcolsep}{.2em} 
\renewcommand{\arraystretch}{1.3}
\caption{Proportion of genes attributed to each putative regulatory class and respective improvement in prediction accuracy $\Delta\mathrm{adj.}R^2$ (relative to a traditional model constructed from the same data) for both H1-hESC and GM12878 cell lines. Adjusted $R^2$ scores were calculated using separate RNA-seq replicates for training and evaluation, as described in Section~\ref{ss:evaluation}.\label{tab:tree1results}}
\begin{center}
\begin{tabular}{c|cc|cc|cc}
\multicolumn{1}{l|}{} & \multicolumn{2}{c|}{\textbf{MMFS$^+$}}                                     & \multicolumn{2}{c|}{\textbf{H2A.Z$^+$}}                                       & \multicolumn{2}{c}{\textbf{H2A.Z$^-$}}                                         \\
                      & \multicolumn{1}{c}{\textbf{genes}} & \multicolumn{1}{c|}{$\Delta\mathrm{adj.}R^2$} & \multicolumn{1}{c}{\textbf{genes}} & \multicolumn{1}{c|}{$\Delta\mathrm{adj.}R^2$} & \multicolumn{1}{c}{\textbf{genes}} & \multicolumn{1}{c}{$\Delta\mathrm{adj.}R^2$} \\ \hline
\textbf{H1-hESC}      & 46\%                                  & +0.03                                  & 25\%                                  & -0.01                                  & 28\%                                   & +0.05                                  \\
\textbf{GM12878}      & 40\%                                  & +0.06                                  & 29\%                                  & -0.13                                  & 30\%                                  & +0.16
\end{tabular}
\end{center}
\end{table}

By separating genes into subsets exhibiting greater regulatory homogeneity, it is evident from Table~\ref{tab:tree1results} that the inferred relationships between regulatory and expression data are significantly strengthened for the majority of genes; \emph{e.g.} $40\%$ of GM12878 genes are classified as MMFS$^+$, and our ability to predict the expression of these genes improves significantly ($\Delta\mathrm{adj.}R^2 = 0.06$, yielding a model with an overall predicted-versus-measured transcript abundance correlation of Pearson's $r > 0.70$).

Reduction in prediction accuracy is constrained to H2A.Z$^+$ genes, which we speculate is due to inherent heterogeneity in H2A.Z-mediated regulation; \emph{i.e.} H2A.Z is known to both maintain H3K4/27me3 bivalency and flank the TSS during transcriptional activation~\cite{raisner2005histone}. It is likely that integrating further histone modifications or related data (\emph{e.g.} DNase-I hypersensitivity) would allow TREEOME to resolve this heterogeneity.

%

\section{Conclusions}


In this study, we have demonstrated that a decision tree-based analytical framework (TREEOME) is able to improve prediction accuracy of regression models for genome-wide RNA transcript abundance by separating genes into putative regulatory classes. We demonstrated the effectiveness of TREEOME by providing the first integration of DNA methylation (bisulfite-seq) and histone modification (ChIP-seq) data to accurately predict genome-wide RNA transcript abundance (RNA-seq) for H1-hESC and GM12878 cell lines.

As described in Section~\ref{sec:introduction}, the utility of predictive gene expression modelling is not the ability to predict RNA levels, but rather the insights into epigenetic regulation of gene expression that can be gained by exploring the relationships inferred from the data. Figure~\ref{fig:treeome_workflow} illustrates one of many possible examples of a predictive modelling workflow, in the context of inferring the unknown regulatory roles of a transcription factor from its position weight matrix.

\begin{figure*}[t]
\begin{center}
    \includegraphics[width=0.8\textwidth]{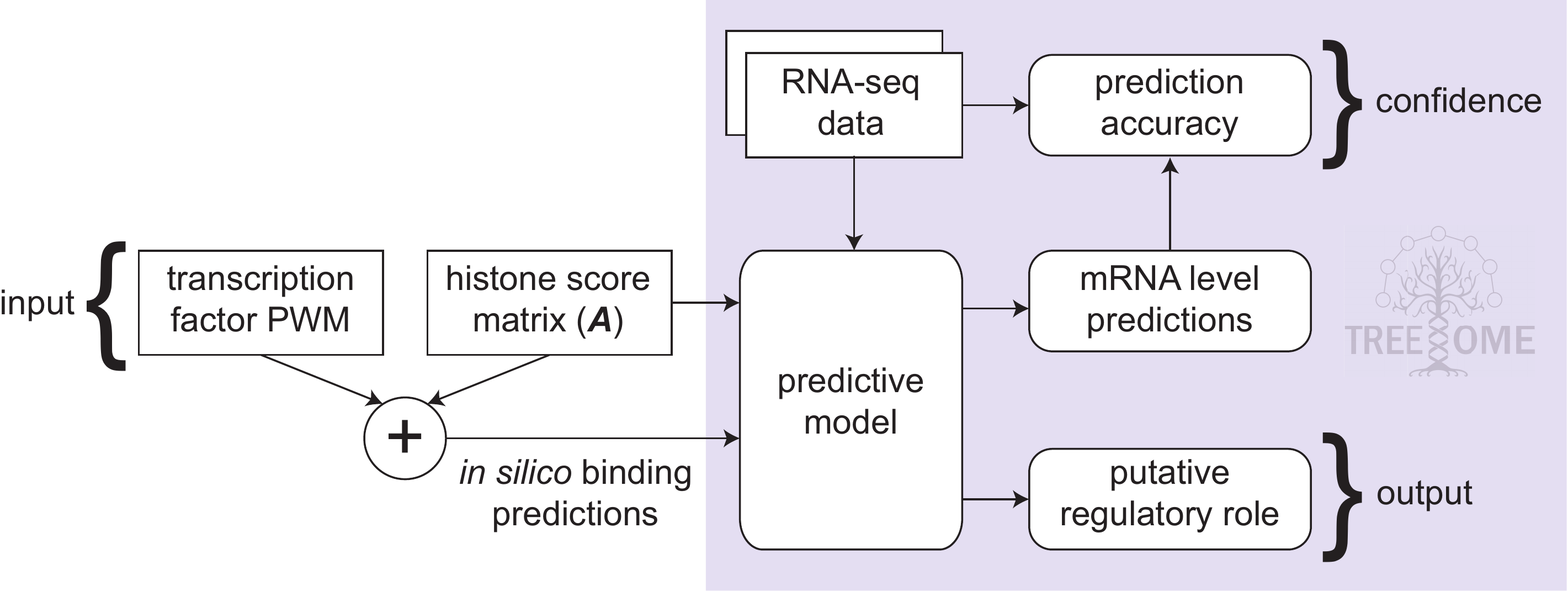}
    \end{center}
    \caption{Example predictive modelling workflow that can be improved by TREEOME integration. Position weight matrices can be combined with an epigenetic prior (\emph{e.g.} H3K4me3 or DNase I hypersensitivity data) to identify putative transcription factor binding sites \emph{in silico} using the Bayesian approach developed by~\cite{cuellar2012epigenetic}. This artificial data can be used to train a model in the same way as actual ChIP-seq data (see Section~\ref{ss:predictive}) to yield models of near-equivalent prediction accuracy~\cite{mcleay2012genome}. The putative regulatory role of the transcription factor can then be derived using PCA, as described in Section~\ref{ss:pca}.}
    \label{fig:treeome_workflow}
\end{figure*}

One limitation of this approach is that it assumes that the functional role of a bound transcription factor is independent of the local chromatin landscape. This is certainly not the case; \emph{e.g.} pioneer transcription factors are able to directly engage nucleosomal DNA~\cite{zaret2011pioneer}, although translating their activating effect to proximal genes requires a subsequent cascade of chromatin-remodelling events that is impossible in the presence of DNA methylation~\cite{jones2012functions}. This limitation is removed by integrating DNA methylation data (via TREEOME) into the blue-highlighted component of Figure~\ref{fig:treeome_workflow} (as described in Section~\ref{ss:decisiontrees}), allowing the derivation of separate regulatory roles and associated confidence values (prediction accuracy) for both methylated and unmethylated genes.

Our TREEOME analysis was prefaced by the first quantitative evaluation of several methods of quantifying gene-level DNA methylation events, which have widespread potential in facilitating future gene-level (rather than CpG-level) differential methylation analyses. We found that the (promoter-localised) mean methylation fraction by site (MMFS) score yields the greatest anti-correlation with gene expression levels in both H1-hESC and GM12878.

We have endeavoured to demonstrate the utility of TREEOME in a practical context and unsupervised manner. The four histone modifications were selected due to their highly conditional and indirect regulatory influence; integrating elements with more direct effects (\emph{e.g.} histone lysine acetylations or DNase I hypersensitivity) would undoubtedly improve prediction accuracy, as demonstrated in previous studies (\emph{e.g.}~\cite{karlic2010histone}. The unsupervised TREEOME threshold selection process could likewise be replaced to capture prior biological knowledge (\emph{e.g.} known methylated genes) or directly-optimised against global prediction accuracy, although we maintain that the latter approach would lose the biological meaning (regulated-or-not) underlying our methodology.

\section*{Acknowledgement}

This work was supported by an Australian Postgraduate Award [D.M.B.]; the Australian Federal and Victoria State Governments and the Australian Research Council through the ICT Centre of Excellence program, National ICT Australia (NICTA) [D.M.B., E.J.C.]; and the Australian Research Council Centre of Excellence in Convergent Bio-Nano Science and Technology (project number CE140100036) [E.J.C.]. The views expressed herein are those of the authors and are not necessarily those of NICTA or the Australian Research Council.

\bibliographystyle{splncs}
\bibliography{budden2015treeome}

\end{document}